\documentclass[11pt]{article}
\usepackage[margin=1in]{geometry}
\usepackage[T1]{fontenc}
\usepackage[utf8]{inputenc}
\usepackage{lmodern}
\usepackage{amsmath,amssymb,amsfonts}
\usepackage{graphicx}
\usepackage{booktabs}
\usepackage{array}
\usepackage{longtable}
\usepackage{tabularx}
\usepackage{multirow}
\usepackage{xcolor}
\usepackage{tikz}
\usetikzlibrary{arrows.meta,positioning,fit,calc,shapes.geometric}
\usepackage{pgfplots}
\pgfplotsset{compat=1.18}
\usepackage[round,authoryear]{natbib}
\usepackage{hyperref}
\hypersetup{colorlinks=true, linkcolor=blue!50!black, citecolor=blue!50!black, urlcolor=blue!50!black}
\usepackage{caption}
\usepackage{enumitem}
\usepackage{float}
\usepackage{pdflscape}
\usepackage{setspace}
\newcolumntype{L}[1]{>{\raggedright\arraybackslash}p{#1}}
\newcolumntype{Y}{>{\raggedright\arraybackslash}X}

\title{Boom, Bubble, or Buildout?\\A Multi-Method Evaluation of Whether Artificial Intelligence Is in an Ongoing Financial Bubble}
\author{Qianan Wang \and Zen Chen}
\date{May 2026}

\begin{document}
\maketitle

\begin{abstract}
The rapid expansion of artificial intelligence (AI) investment has revived a recurrent question in financial economics: are AI-related assets experiencing a bubble, or is the market capitalizing a durable general-purpose technology? This paper develops a hybrid review and diagnostic framework for evaluating whether AI is in an ongoing financial bubble as of May 2026. The analysis begins from asset-pricing foundations in state prices, stochastic discount factors, martingale valuation, and pricing kernels, then connects these foundations to rational bubbles, behavioral bubbles, technology manias, and modern econometric bubble-detection methods. The paper argues that the AI-bubble question should not be reduced to a binary claim about all AI-related assets. Instead, it should be analyzed as a segmented problem across the AI stack: semiconductors, cloud infrastructure, data centers and power, foundation models, application software, agentic AI, and venture-backed startups. Current evidence shows both genuine fundamentals and bubble-like fragilities. On the fundamental side, realized revenue growth, enterprise adoption, and productivity evidence support a nontrivial share of AI valuations. On the fragile side, capital expenditure has accelerated faster than observed monetization in some layers, private-market valuations are concentrated in a small number of firms, and investor narratives often capitalize future productivity gains before they have appeared in cash flows. The paper proposes a five-pillar diagnostic framework that combines fundamental valuation, residual-exuberance tests, SADF/GSADF explosive-root procedures, LPPL/HLPPL price-pattern diagnostics, sentiment and issuance measures, and capex-payback analysis. The central conclusion is that AI is best understood as a real technological revolution with localized bubble dynamics rather than as either a pure speculative mania or a bubble-free productivity miracle.
\end{abstract}


\tableofcontents
\newpage

\section{Introduction}

Artificial intelligence has become the dominant investment narrative in global equity, venture-capital, cloud-computing, semiconductor, and data-center markets. The 2023--2026 period has combined four features that often appear in earlier episodes of technological exuberance: a plausible general-purpose technology, sharp changes in expected cash flows, concentrated financial winners, and uncertainty about how quickly real productivity gains will diffuse through the economy. These features make AI a difficult object for bubble diagnosis. A technology can be economically transformative and still be overcapitalized during its installation phase; conversely, prices can look expensive under conventional ratios because investors are capitalizing legitimate growth options under uncertainty.

The empirical tension is unusually stark. AI investment and adoption have accelerated rapidly. The 2026 AI Index reports that U.S. private AI investment reached \$285.9 billion in 2025 and that generative AI investment became a major driver of global corporate AI spending \citep{stanford2026aiindex}. Energy infrastructure has become a binding part of the AI investment story: the International Energy Agency projects that data-center electricity consumption could roughly double from 485 TWh in 2025 to 950 TWh by 2030, with AI-focused data centers growing faster than the broader data-center category \citep{iea2026energyai}. Capital-market estimates are similarly large. Goldman Sachs' 2026 supply-side model implies AI-related annual capital expenditure of about \$765 billion in 2026, rising to about \$1.6 trillion by 2031 under baseline assumptions \citep{goldman2026tracking}. At the firm level, the leading AI-infrastructure supplier NVIDIA reported record first-quarter fiscal 2027 revenue of \$81.6 billion, including \$75.2 billion in data-center revenue \citep{nvidia2026q1}. These facts complicate claims that the AI boom is mere fantasy.

At the same time, scale alone does not prove sustainability. The capitalized value of a technology boom depends not only on how much firms invest, but on whether that investment converts into durable cash flows after depreciation, energy costs, competition, and customer adoption. The dot-com episode is the canonical reminder that real technology and inflated valuation can coexist. Internet infrastructure, search, e-commerce, and cloud computing ultimately became central to the economy, yet many public internet firms of the late 1990s were dramatically overvalued relative to eventual cash flows \citep{ofek2003,greenwood2019}. A similar distinction is necessary for AI. The question is not whether AI matters. It is whether prices in specific AI-exposed segments have detached from plausible fundamentals.

This paper makes three contributions. First, it integrates bubble theory with asset-pricing foundations. Definitions of bubbles often become rhetorical because they lack a valuation benchmark. A price can be high because expected cash flows are high, discount rates are low, growth options are valuable, or resale expectations are self-referential. The stochastic-discount-factor equation provides a disciplined starting point: a price is supported by fundamentals when it can be represented as the discounted value of future payoffs under an economically interpretable pricing kernel. A bubble component is the part of price that survives after such fundamentals have been modeled. This framing connects classical state-price theory, martingale valuation, and modern learned pricing kernels \citep{arrow1953,debreu1959,harrison1979,harrison1981,hansen1991,cochrane2005,chen2026probmeasures}.

Second, the paper treats AI as a segmented financial ecosystem rather than a monolithic asset class. The AI stack includes hardware, cloud platforms, data centers, power procurement, model developers, applications, agents, and enterprise users. Bubble risk can arise in one layer even when another layer is fundamentally supported. Semiconductor leaders may generate realized cash flows, while application startups face commoditization; hyperscalers may have strong balance sheets, while data-center developers may face power bottlenecks and useful-life uncertainty; foundation-model firms may grow revenue rapidly, while private valuations may embed circular financing and aggressive margin assumptions. A useful diagnosis must therefore map valuation, adoption, and capital intensity by layer.

Third, the paper proposes a multi-method diagnostic framework. A single test cannot settle whether AI is a bubble. Discounted cash-flow models are sensitive to assumptions; econometric tests can identify explosive prices but not prove irrationality; LPPL models can detect bubble-like acceleration but require careful calibration; sentiment measures can identify hype but not value; and capex-payback analysis depends on accounting for useful lives, utilization, and customer willingness to pay. The proposed framework combines these tools into a segment-level evidence matrix. It extends standard bubble diagnostics by incorporating machine-learning-based asset-pricing ideas and hype-augmented LPPL methods, including the HLPPL framework of \citet{cao2025hlppl} and the broader evolution of machine learning in finance synthesized by \citet{chen2026mlfinance}.

The central argument is deliberately balanced. As of May 2026, AI markets display bubble-like characteristics in some segments: highly valued private foundation-model firms, thin-profit AI application firms, speculative data-center projects, and assets whose valuations depend mainly on distant productivity gains. Yet the largest public AI infrastructure firms also have substantial fundamental support through realized revenue, high margins, and bottleneck rents. The best characterization is therefore a \emph{real technological revolution with localized bubble dynamics}. This conclusion is consistent with models of technological revolutions in which prices of innovative firms can exhibit bubble-like patterns during periods of uncertainty and learning \citep{pastor2009}, and with empirical evidence that sharp industry run-ups predict higher crash probabilities but not necessarily lower average returns \citep{greenwood2019}.

The rest of the paper proceeds as follows. Section \ref{sec:definitions} defines bubbles and explains why the AI case requires a segmented approach. Section \ref{sec:assetpricing} reviews asset-pricing foundations. Section \ref{sec:history} compares AI with prior technology booms. Section \ref{sec:econ} reviews AI productivity, adoption, and macroeconomic evidence. Section \ref{sec:stack} maps bubble risk across the AI stack. Section \ref{sec:methods} reviews econometric, LPPL, and machine-learning approaches to bubble detection. Section \ref{sec:framework} develops the proposed diagnostic framework. Section \ref{sec:empirical} presents an empirical design. Section \ref{sec:evidence} summarizes current evidence for and against an AI bubble. Section \ref{sec:policy} discusses financial-stability and governance implications. Section \ref{sec:limitations} states limitations and open problems. Section \ref{sec:conclusion} concludes.

\section{What Counts as a Bubble?}\label{sec:definitions}

The term \emph{bubble} is widely used but conceptually unstable. In casual market commentary, a bubble often means simply that prices have risen quickly. In financial economics, the concept is narrower. A bubble is a component of price that cannot be justified by expected future cash flows under a plausible discounting model, or a price process in which valuation is sustained largely by expectations of resale to future buyers. \citet{stiglitz1990} states the core intuition: if the reason a price is high today is that investors expect it to be even higher tomorrow rather than because of fundamentals, the asset has a bubble component. This definition immediately raises the benchmark problem. To say that price exceeds fundamentals, one must define fundamentals.

Rational-bubble theory shows that price can contain a bubble component without violating investors' expectations. In a representative equation, the observed price $P_t$ equals a fundamental component $F_t$ plus a bubble component $B_t$:
\begin{equation}
P_t = F_t + B_t.
\end{equation}
The bubble component satisfies an intertemporal pricing condition,
\begin{equation}
B_t = E_t\left[M_{t+1}B_{t+1}\right],
\end{equation}
where $M_{t+1}$ is a stochastic discount factor. In a risk-neutral, constant-discount-rate special case, this condition implies that the bubble must grow at the required rate of return as long as it survives. This does not make the bubble socially efficient or empirically easy to identify; it means that bubble prices can be internally consistent in a model with expectations, constraints, and resale options \citep{blanchard1982,tirole1985,santos1997}.

Behavioral and limits-to-arbitrage theories emphasize a different mechanism. Bubbles can persist because pessimists are constrained, arbitrage is risky, investors disagree, sentiment affects demand, and sophisticated traders may ride a bubble rather than eliminate it. Classic models of heterogeneous beliefs and short-sale constraints show how optimists can set prices when pessimists cannot fully express negative views \citep{miller1977,harrison1978}. Limits to arbitrage make mispricing difficult to correct even when some traders recognize it \citep{shleifer1997}. Models of investor sentiment, overconfidence, and disagreement explain why beliefs may extrapolate from recent returns and why bubbles can become fragile when beliefs reverse \citep{barberis1998,shiller2000,scheinkman2003,abreu2003,hong2006,greenwood2014,brunnermeier2008}.

Crisis narratives add credit, leverage, and institutional feedback. \citet{kindleberger1978} and \citet{minsky1992} interpret bubbles as part of broader financial cycles in which displacement, credit expansion, speculative finance, and eventual distress interact, while models of bubbles and crises show how intermediated credit and asset-price feedback can amplify the cycle \citep{allen2000}. These narratives are especially relevant for AI if capital expenditure becomes increasingly debt-financed, if private funding rounds become circular, or if collateral values depend on the continued appreciation of AI-related assets. The AI boom has not yet displayed the broad household leverage of housing bubbles, but it has displayed capital-intensity and concentration features that make financial-stability analysis relevant.

Econometric bubble detection operationalizes the concept through price dynamics. Recursive right-tailed unit-root tests identify episodes of explosive behavior in asset prices \citep{phillips2011,phillips2015a,phillips2015b}. LPPL methods identify faster-than-exponential growth decorated by accelerating oscillations as a signature of positive feedback and critical transition risk \citep{johansen2000,sornette2001,sornette2003,filimonov2013}. More recent methods add textual hype and machine-learning classifiers to price-based signals \citep{cao2025hlppl}. These tools do not establish a full welfare-theoretic definition of a bubble, but they can identify the kinds of instability that matter for investors and policymakers.

Table \ref{tab:definitions} summarizes the major definitions used in this paper. The key lesson is that no definition is sufficient alone. A useful AI-bubble diagnosis should combine valuation, expectation formation, market constraints, price dynamics, and capital-allocation consequences.

\begin{table}[H]
\centering
\caption{Definitions of financial bubbles and their relevance to AI}\label{tab:definitions}
\small
\begin{tabularx}{\textwidth}{L{3.0cm}Y Y}
\toprule
Definition & Core idea & Relevance to AI markets \\
\midrule
Fundamental-value bubble & Price exceeds discounted expected cash flows under a plausible pricing kernel. & Requires cash-flow scenarios for chips, cloud, models, applications, and data centers. \\
Rational bubble & A self-consistent price component grows because investors expect future resale value. & Useful for analyzing firms whose valuations rely on future capital-market access rather than current profits. \\
Behavioral bubble & Sentiment, extrapolation, overconfidence, and disagreement raise prices above fundamentals. & Applicable to AI narratives, retail attention, private-market competition, and thematic ETFs. \\
Minsky-Kindleberger bubble & Credit expansion and speculative finance amplify a technological displacement. & Relevant if AI capex becomes debt-heavy or circular financing becomes material. \\
Econometric bubble & Prices display explosive dynamics, date-stamped by recursive tests or LPPL patterns. & Enables real-time testing of AI-exposed public equities and sector baskets. \\
Segmented bubble & Bubble-like pricing is concentrated in some layers while others remain fundamentally supported. & The most plausible current characterization of AI-related markets. \\
\bottomrule
\end{tabularx}
\end{table}

\section{Asset-Pricing Foundations for Bubble Diagnosis}\label{sec:assetpricing}

A claim that AI is in a bubble must be grounded in an asset-pricing model. The conceptual lineage begins with Bachelier's probabilistic model of speculation and Fisher's intertemporal theory of interest \citep{bachelier1900,fisher1930}. Arrow-Debreu state-contingent claims formalized the idea that securities can be priced by state prices in complete markets \citep{arrow1953,debreu1959}. Samuelson and Fama then connected competitive information processing to unpredictable price changes and efficient-market benchmarks \citep{samuelson1965,fama1970}. Arbitrage pricing and option-implied state-price ideas added further links between observed securities and latent valuation measures \citep{ross1976,breeden1978}. Grossman and Stiglitz complicated the strongest version of efficiency by showing that perfectly informative prices would undermine the incentive to acquire information \citep{grossman1980}. Shiller's variance-bound critique further suggested that prices can move more than subsequent dividends appear to justify \citep{shiller1981}.

Modern asset pricing expresses valuation through the stochastic discount factor (SDF):
\begin{equation}\label{eq:sdf}
P_{i,t} = E_t\left[M_{t+1}X_{i,t+1}\right],
\end{equation}
where $P_{i,t}$ is the price of asset $i$, $X_{i,t+1}$ is its payoff, and $M_{t+1}$ prices risk across states. In a one-period complete market, state prices are equivalent to risk-adjusted probabilities multiplied by discount factors. In continuous time, the Black-Scholes-Merton framework and martingale valuation theory use equivalent martingale measures to express prices as discounted expectations under risk-neutral probabilities \citep{black1973,merton1973,harrison1979,harrison1981}. The fundamental theorem of asset pricing links absence of arbitrage to the existence of equivalent martingale measures under appropriate conditions \citep{delbaen1994,duffie2001}. The Hansen-Jagannathan bounds show how asset returns restrict admissible SDFs \citep{hansen1991}; Cochrane's synthesis places the SDF at the center of empirical asset pricing \citep{cochrane2005}.

For AI valuation, the SDF framework clarifies three issues. First, high expected cash-flow growth can justify high prices if the growth is likely, persistent, and appropriately discounted. Second, risk matters: AI cash flows may be highly systematic if they are tied to the investment cycle, energy prices, concentration in mega-cap technology firms, or macroeconomic productivity expectations. Third, fundamentals are not a single number. They are a distribution of payoffs across states, with state prices that depend on marginal utility, risk aversion, constraints, and uncertainty.

\citet{chen2026probmeasures} provides a useful historical synthesis for this paper because it interprets asset-pricing theory as the development of probability measures for valuation. The progression from state prices to martingale measures, forward measures, stochastic discount factors, and learned pricing kernels is especially relevant to AI. If modern markets process information through text, attention, and sentiment, then the valuation measure is not merely a historical-frequency distribution. It is an information-adjusted measure shaped by investor beliefs, risk premia, liquidity, and machine-readable narratives. This does not make prices correct; it means that diagnosing mispricing requires distinguishing genuine information transformation from hype-driven probability distortion.

The AI-bubble question can therefore be expressed as a residual valuation problem. Let $F_{i,t}$ denote a model-implied fundamental value for an AI-exposed asset, estimated from cash-flow scenarios and discount rates. Define residual exuberance as
\begin{equation}\label{eq:residexuberance}
RE_{i,t}=\log P_{i,t} - \log \widehat{F}_{i,t}.
\end{equation}
A high residual does not automatically prove a bubble, because $\widehat{F}_{i,t}$ may omit growth options, intangibles, or network effects. But a persistent and rising residual, combined with explosive price dynamics, sentiment extremes, and weak monetization, is stronger evidence of bubble risk than any indicator alone.

In practice, the fundamental value $F_{i,t}$ can be estimated using discounted free cash flow, residual income, or SDF-based valuation. A stylized DCF representation is
\begin{equation}\label{eq:dcf}
V_0 = \sum_{t=1}^{T}\frac{FCF_t}{(1+r)^t}+\frac{TV_T}{(1+r)^T},
\end{equation}
where $FCF_t$ is free cash flow, $r$ is a risk-adjusted discount rate, and $TV_T$ is terminal value. In AI infrastructure, $FCF_t$ must account for capex, depreciation, power, networking, cooling, and customer concentration. In model and application firms, it must account for inference costs, competition, churn, and the durability of pricing power. The same headline revenue growth can imply very different valuations depending on these assumptions.

This asset-pricing foundation also explains why efficient-market skepticism about bubbles remains important. \citet{fama2014} argues that asset-pricing tests are joint tests of market efficiency and the model of expected returns. The AI-bubble diagnosis inherits this joint-test problem. If an AI stock looks expensive, one explanation is mispricing; another is that the analyst's model omits growth options or misprices risk. The framework in this paper does not claim to escape the joint-test problem. It reduces the risk of rhetorical overclaiming by requiring multiple forms of evidence before labeling a segment a bubble.

\section{Technology Revolutions and Market Overcapitalization}\label{sec:history}

Financial bubbles often accompany technological revolutions because new technologies create genuine option value while making valuation unusually uncertain. Schumpeterian innovation reallocates capital toward new combinations, new firms, and new infrastructures \citep{schumpeter1942}. Perez's installation-deployment framework argues that technological revolutions often pass through a financially exuberant installation phase before productivity gains are fully integrated into production systems \citep{perez2002}. Historical analogies include railroads, electricity, radio, automobiles, personal computers, telecommunications, the internet, and cryptoassets.

The analogy to prior technology booms is useful but dangerous. AI differs from railroads, dot-com firms, and crypto in its supply chain, monetization model, and institutional embedding. Yet the general pattern is familiar. A plausible general-purpose technology induces large complementary investments before realized productivity data can verify the social return. \citet{david1990} used the electric dynamo and computer to show that productivity gains can be delayed by organizational complements. \citet{jovanovic2005} describe general-purpose technologies as pervasive technologies that improve over time and generate complementary innovation. \citet{brynjolfsson2021} formalize the productivity J-curve: early investment in intangible complements can depress measured productivity before later gains appear. These models help explain why early AI productivity data may understate long-run value, but they also imply a risk of overinvestment during the installation phase.

\citet{pastor2009} provide a direct bridge from technology revolutions to stock prices. In their model, innovative-firm valuations can look bubble-like when investors learn about the productivity of a new technology. Uncertainty shifts from idiosyncratic to systematic as the technology diffuses, and stock prices can surge before later correction. This mechanism fits AI better than a simple irrational-mania story because it allows real technological value and high valuation uncertainty to coexist.

The dot-com episode remains the closest public-equity analogy. Internet technology was transformative, but many internet stocks in the late 1990s were priced beyond plausible cash flows. \citet{ofek2003} emphasize heterogeneous beliefs and short-sale constraints in the rise and fall of internet stocks. \citet{greenwood2019} evaluate industry run-ups and find that sharp price increases do not necessarily predict low average returns, but they do predict a higher probability of a crash, especially when accompanied by volatility, issuance, and other run-up characteristics. This distinction is central for AI. A large price increase in an AI-exposed industry does not prove that the average return must be negative. It may indicate that the distribution has become more crash-prone.

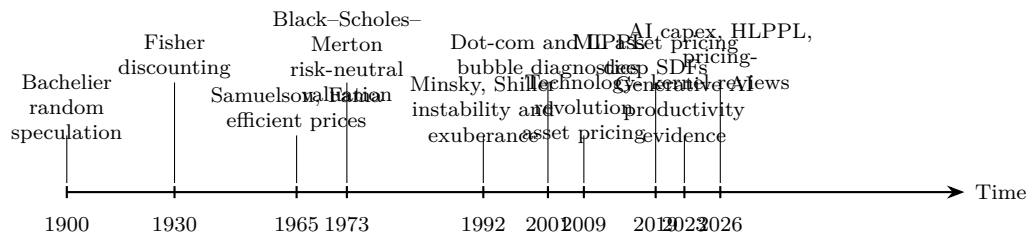
\begin{figure}[H]
\centering
\begin{tikzpicture}[x=0.095cm,y=1cm,>=Stealth, every node/.style={font=\scriptsize}]
  \draw[->, thick] (0,0) -- (125,0) node[right]{Time};
  \foreach \x/\yr in {0/1900,15/1930,32/1965,39/1973,58/1992,67/2001,72/2009,82/2019,86/2023,91/2026}
    \draw[thick] (\x,0.08) -- (\x,-0.08) node[below=3pt]{\yr};
  \node[align=center, text width=2.2cm] at (0,1.1) {Bachelier\\random speculation};
  \draw (0,0.12) -- (0,0.75);
  \node[align=center, text width=2.2cm] at (15,1.8) {Fisher\\discounting};
  \draw (15,0.12) -- (15,1.45);
  \node[align=center, text width=2.5cm] at (32,1.1) {Samuelson, Fama\\efficient prices};
  \draw (32,0.12) -- (32,0.75);
  \node[align=center, text width=2.7cm] at (39,1.8) {Black--Scholes--Merton\\risk-neutral valuation};
  \draw (39,0.12) -- (39,1.45);
  \node[align=center, text width=2.4cm] at (58,1.1) {Minsky, Shiller\\instability and exuberance};
  \draw (58,0.12) -- (58,0.75);
  \node[align=center, text width=2.7cm] at (67,1.8) {Dot-com and LPPL\\bubble diagnostics};
  \draw (67,0.12) -- (67,1.45);
  \node[align=center, text width=2.7cm] at (72,1.1) {Technology-revolution\\asset pricing};
  \draw (72,0.12) -- (72,0.75);
  \node[align=center, text width=2.6cm] at (82,1.8) {ML asset pricing\\deep SDFs};
  \draw (82,0.12) -- (82,1.45);
  \node[align=center, text width=2.6cm] at (86,1.1) {Generative AI\\productivity evidence};
  \draw (86,0.12) -- (86,0.75);
  \node[align=center, text width=2.8cm] at (91,1.8) {AI capex, HLPPL,\\pricing-kernel reviews};
  \draw (91,0.12) -- (91,1.45);
\end{tikzpicture}
\caption{Conceptual timeline connecting asset-pricing foundations, bubble diagnostics, machine learning in finance, and the 2023--2026 AI boom. The figure is a conceptual synthesis rather than a data plot.}\label{fig:timeline}
\end{figure}

Table \ref{tab:analogies} compares AI with historical technology booms. The comparison supports two conclusions. First, technological revolutions often produce durable winners and large deadweight losses simultaneously. Second, the relevant bubble question is not whether the technology survives, but whether capital is allocated at prices that future cash flows can justify.

\begin{table}[H]
\centering
\caption{Historical technology booms and lessons for AI}\label{tab:analogies}
\small
\begin{tabularx}{\textwidth}{L{2.2cm}L{2.8cm}Y Y}
\toprule
Episode & Fundamental innovation & Bubble mechanism & AI lesson \\
\midrule
Railroads & Transport network, market integration & Overbuilding, leverage, land speculation & Infrastructure can be socially valuable even when private investors overpay. \\
Electricity & General-purpose energy platform & Complementary investments lagged technology & Productivity gains may arrive after organizational redesign. \\
Radio and electronics & New communication and consumer markets & Narrative valuation and uncertain business models & Consumer excitement can outrun monetization. \\
Dot-com & Internet protocols, e-commerce, search & Heterogeneous beliefs, IPO wave, weak profits & Real technology did not prevent widespread overvaluation. \\
Telecom fiber & Internet backbone capacity & Capex boom, debt, overcapacity & AI data-center and networking capex may face similar utilization risk. \\
Cloud computing & Scalable enterprise infrastructure & Platform concentration but durable cash flows & Infrastructure leaders can be fundamentally supported. \\
Crypto & Decentralized ledgers, speculative tokens & Reflexive narratives and token issuance & Not all technological narratives produce cash-flow assets. \\
Artificial intelligence & Prediction, generation, reasoning, automation & Capex extrapolation, private valuation, hype, circular demand & Bubble risk likely differs across stack layers. \\
\bottomrule
\end{tabularx}
\end{table}

\section{AI as a General-Purpose Technology: Productivity, Diffusion, and Lags}\label{sec:econ}

The strongest argument against a pure AI-bubble diagnosis is that AI has already produced measurable productivity effects in controlled settings and is plausibly a general-purpose technology. The economics of AI begins with prediction. \citet{agrawal2018} argue that AI lowers the cost of prediction, changing the value of judgment, data, and complements. The broader research agenda in \citet{agrawal2019} situates AI within labor markets, industrial organization, and productivity. \citet{furman2019} review early evidence that AI can reshape the economy through automation, complementarity, competition, and policy channels.

Empirical evidence on generative AI shows positive but heterogeneous effects. \citet{noy2023} find that ChatGPT reduced task completion time and improved output quality in professional writing tasks. \citet{brynjolfsson2025} study a deployed generative AI assistant in customer support and find an average productivity gain, with larger gains for less experienced workers. \citet{dellacqua2023} introduce the jagged frontier: AI improves performance on tasks inside its capability frontier but can worsen performance on tasks outside it. These studies support the view that AI can create real economic value, but they do not directly imply that market valuations are correct. Task-level productivity does not automatically scale into aggregate profits.

Macroeconomic models are more cautious. \citet{acemoglu2024} argues that the aggregate effect of AI depends on the share of tasks affected, cost savings, new task creation, and distributional channels. AI could be significant without generating the enormous productivity acceleration implied by the most optimistic market narratives. Labor-exposure measures likewise suggest that many tasks may be affected, but exposure is not adoption, adoption is not productivity, and productivity is not necessarily captured by public shareholders \citep{acemoglu2018,felten2021,eloundou2023}. The IMF's 2026 scenario-planning note frames AI as a macro-critical transition whose outcomes depend on diffusion speed, institutional readiness, financial stability, and global coordination \citep{imf2026ai}.

Enterprise adoption data reinforce this ambiguity. McKinsey's 2025 survey reports broad AI use but persistent difficulties moving from pilots to scaled value \citep{mckinsey2025}. PwC's 2026 AI performance study emphasizes that AI-driven returns are concentrated among firms with strong complementary capabilities \citep{pwc2026}. These findings fit the productivity J-curve: large early investments may be necessary, but returns depend on intangible complements such as data quality, workflow redesign, governance, talent, and organizational learning \citep{brynjolfsson2021}. Financial markets may be capitalizing the eventual right side of the J-curve before the economy has traversed the left side.

This productivity evidence has two implications for bubble diagnosis. First, it weakens simplistic claims that AI valuations are unsupported because AI lacks real use cases. There are real use cases. Second, it strengthens the need for segmentation. If AI value accrues primarily to firms that control compute, distribution, data, and workflow integration, then application-layer firms without durable moats may be overvalued even if AI as a technology succeeds. Similarly, if productivity gains accrue to customers rather than suppliers, supplier valuations may overstate the portion of surplus captured by shareholders.

\section{The AI Stack and Segmented Bubble Risk}\label{sec:stack}

\begin{figure}[H]
\centering
\begin{tikzpicture}[node distance=0.55cm, every node/.style={font=\scriptsize}, box/.style={draw, rounded corners, align=center, minimum width=3.1cm, minimum height=0.75cm, fill=gray!8}, risk/.style={draw, rounded corners, align=left, text width=5.2cm, fill=gray!4}]
  \node[box] (chips) {Semiconductors\\GPUs, ASICs, memory};
  \node[box, below=of chips] (cloud) {Cloud and networking\\training and inference clusters};
  \node[box, below=of cloud] (data) {Data centers and power\\land, grid, cooling, PPAs};
  \node[box, below=of data] (models) {Foundation models\\frontier labs and open models};
  \node[box, below=of models] (apps) {Applications and agents\\workflow products, copilots};
  \node[box, below=of apps] (users) {Enterprise and consumer adoption\\revenue, cost savings, productivity};
  \foreach \a/\b in {chips/cloud,cloud/data,data/models,models/apps,apps/users} \draw[->, thick] (\a) -- (\b);
  \node[risk, right=1.1cm of chips] (r1) {\textbf{Fundamental support:} scarcity rents, realized demand, high gross margins.\\\textbf{Bubble risk:} extrapolated supply shortage, customer concentration, geopolitical supply chains.};
  \node[risk, right=1.1cm of data] (r2) {\textbf{Fundamental support:} durable infrastructure and energy constraints.\\\textbf{Bubble risk:} overbuilding, short chip lives, grid bottlenecks, debt-financed capacity.};
  \node[risk, right=1.1cm of apps] (r3) {\textbf{Fundamental support:} workflow automation and measurable productivity.\\\textbf{Bubble risk:} low switching costs, model commoditization, thin margins, narrative valuation.};
  \draw[->, dashed] (chips.east) -- (r1.west);
  \draw[->, dashed] (data.east) -- (r2.west);
  \draw[->, dashed] (apps.east) -- (r3.west);
\end{tikzpicture}
\caption{Segmented AI stack and bubble-risk channels. The diagram is a conceptual synthesis showing why AI bubble diagnosis should be performed by layer rather than at the level of a single AI theme.}\label{fig:stack}
\end{figure}
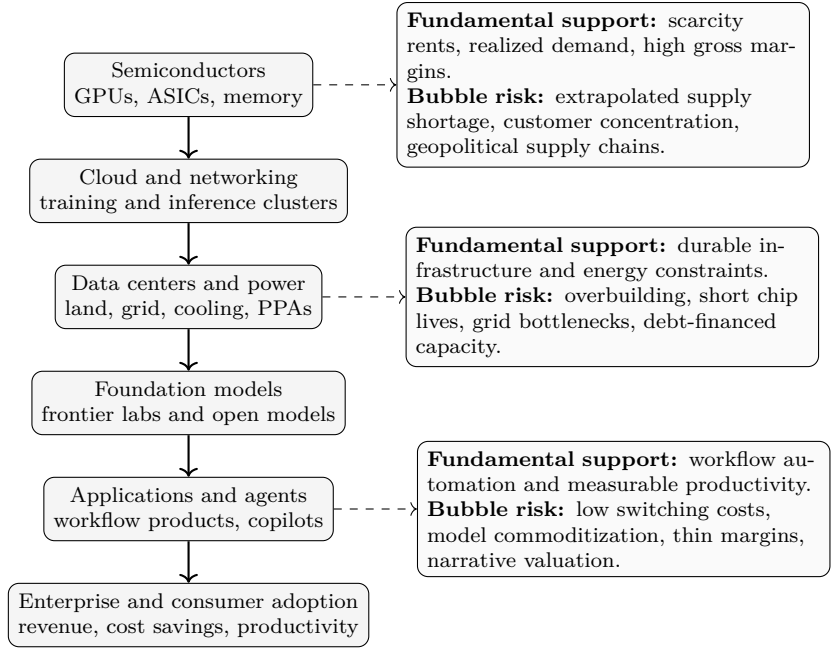

AI markets should be decomposed into layers because each layer has different economics. Figure \ref{fig:stack} maps the stack and the main risk channels. The upstream layers--chips, memory, networking, cloud clusters, and data centers--are capital intensive and supply constrained. The downstream layers--models, applications, agents, and enterprise workflows--are more exposed to competition, adoption friction, and monetization uncertainty.

\subsection{Semiconductors and bottleneck rents}

AI accelerators, high-bandwidth memory, advanced packaging, and networking equipment have generated the most visible realized revenues. These markets can support high valuations when demand exceeds supply and when switching costs, software ecosystems, and manufacturing constraints generate economic rents. The case against a bubble in the leading semiconductor firms is therefore stronger than in many downstream startups: cash flows exist, margins are high, and customers are identifiable.

Yet semiconductor valuations still carry bubble risk. First, demand is concentrated among a small number of hyperscalers and frontier-model firms. Second, customers may double order during shortages, producing temporary revenue that reverses when supply normalizes. Third, chip useful lives may shorten if model architectures change or if inference shifts toward more efficient hardware. Fourth, margins may compress if competition from custom silicon, alternative accelerators, or open hardware ecosystems increases. A semiconductor leader can be fundamentally strong and still overpriced if markets extrapolate scarcity rents indefinitely.

\subsection{Cloud platforms and hyperscaler capex}

Cloud platforms occupy a middle position. They have durable enterprise relationships, large balance sheets, and the ability to monetize AI through cloud services, software bundles, and internal productivity. Their capex is partly defensive: not investing may risk losing enterprise workloads and model partnerships. But defensive capex can still destroy value if the resulting capacity earns low returns on invested capital. The market must therefore evaluate whether AI capex creates incremental cash flow or merely preserves competitive position.

For hyperscalers, the central metric is not gross capex but risk-adjusted return on AI invested capital. A simplified payback condition is
\begin{equation}\label{eq:capexpayback}
\text{AI Revenue}_t \geq \frac{\text{CapEx}_t}{L}+\text{Opex}_t+rK_t,
\end{equation}
where $L$ is the useful life of AI infrastructure, $\text{Opex}_t$ includes power, cooling, networking, labor, and maintenance, $K_t$ is invested capital, and $r$ is the cost of capital. Bubble risk rises if prices assume long useful lives, high utilization, and durable pricing power while actual infrastructure depreciates faster or faces lower utilization.

\subsection{Data centers, power, and physical constraints}

Data centers and power assets are increasingly central to the AI-bubble debate. AI is not only a software boom; it is also an industrial buildout. IEA projections show that data-center electricity demand may become a meaningful share of global electricity growth by 2030 \citep{iea2026energyai}. This creates opportunities for data-center developers, utilities, grid equipment providers, and power producers. It also creates bottlenecks: interconnection queues, permitting, water use, land availability, and local political opposition.

Data-center assets can become bubble-prone when long-duration capital is committed against short-duration technology assumptions. If a data center is financed on the assumption of high GPU utilization for many years, but chip generations change quickly or inference demand migrates geographically, the asset's economic life may be shorter than its financing life. This resembles telecom fiber overbuild in the early 2000s: the infrastructure was eventually useful, but the capital structure and timing were often wrong.

\subsection{Foundation models}

Foundation-model firms combine high growth, strategic importance, and unusually uncertain unit economics. They may capture value through API usage, enterprise subscriptions, coding agents, consumer products, advertising, or model licensing. They may also face high inference costs, model commoditization, safety and governance costs, and competition from open models. Private valuations can be especially difficult to assess because terms, revenue quality, compute commitments, and related-party transactions are often opaque.

For foundation-model firms, bubble risk is linked to the gap between revenue scale and capital requirements. A firm growing annualized revenue rapidly may still be overvalued if gross margins collapse under inference costs, if customer retention is weak, or if model performance becomes commoditized. Conversely, a firm with durable distribution and high-value workflow integration could justify high valuations. The appropriate diagnosis is not to label all foundation-model firms bubbles, but to test whether valuation assumptions imply implausible market share, margins, or capex efficiency.

\subsection{Applications and agentic AI}

The application layer is likely to contain the widest distribution of outcomes. Some firms will embed AI into defensible workflows and capture value. Others will be wrappers around frontier models with weak switching costs. Agentic AI adds a further layer of uncertainty: if agents can plan, use tools, execute workflows, and coordinate with humans, they may create substantial productivity gains. But reliability, auditability, liability, security, and governance constraints may slow adoption in regulated industries.

Machine learning in finance illustrates both the potential and the caution. Early neural-network forecasting in finance dates back to \citet{white1988}, while derivative pricing via learning networks appeared in \citet{hutchinson1994}. Modern empirical asset pricing uses machine learning to model nonlinear predictor interactions \citep{gu2020}, no-arbitrage deep learning \citep{chen2024deeplearning}, and high-dimensional market microstructure patterns \citep{sirignano2019}; broader introductions to deep learning in finance emphasize both representation power and implementation risk \citep{heaton2017}. Yet the literature also emphasizes data snooping, nonstationarity, interpretability, and implementation frictions \citep{dixon2020,lopezdeprado2018}. \citet{chen2026mlfinance} places these developments in a longer arc from early neural-network forecasts to agentic AI, highlighting that financial applications require evidence, validation, governance, and robustness rather than technological novelty alone.

\begin{table}[H]
\centering
\caption{AI stack risk matrix}\label{tab:stackmatrix}
\small
\begin{tabularx}{\textwidth}{L{2.5cm}Y Y Y}
\toprule
Layer & Fundamental support & Bubble-risk mechanism & Diagnostic focus \\
\midrule
Semiconductors & Realized demand, high margins, scarcity rents & Extrapolated shortages, customer concentration, custom-silicon competition & Revenue durability, backlog quality, customer concentration, margin sensitivity \\
Cloud platforms & Enterprise relationships, distribution, balance sheets & Defensive overinvestment and low incremental ROIC & AI revenue attribution, utilization, return on invested capital \\
Data centers and power & Physical bottlenecks, long-term demand for compute & Overbuilding, grid bottlenecks, asset-life mismatch & Contract quality, useful life, power costs, financing structure \\
Foundation models & Rapid adoption, platform potential, strategic value & High training/inference costs, commoditization, opaque private terms & Gross margin after compute, retention, pricing power, related-party revenue \\
Applications & Workflow automation and vertical specialization & Low switching costs, model wrappers, crowded markets & Net revenue retention, workflow depth, customer surplus capture \\
Agentic AI & Potential for end-to-end automation & Reliability, governance, liability, security failures & Task success rates, auditability, human oversight, regulated adoption \\
Venture AI & Optionality and experimentation & Valuation without monetization, financing competition & Down-round risk, revenue quality, capital intensity, exit assumptions \\
\bottomrule
\end{tabularx}
\end{table}

\section{Bubble-Detection Methods}\label{sec:methods}

Bubble diagnosis requires multiple methods because each method identifies a different symptom. This section reviews four major families: fundamentals-based residuals, explosive-root tests, LPPL and HLPPL models, and machine-learning/sentiment approaches.

\subsection{Fundamental valuation and residual exuberance}

Fundamental valuation compares price with expected discounted cash flows. Its strength is interpretability: it forces the analyst to specify revenue, margins, capex, discount rates, and terminal value. Its weakness is sensitivity to assumptions. In AI, small changes in terminal margin, useful life, or market share can dramatically change value. A foundation-model firm may look cheap if it captures a large share of global knowledge-work surplus and expensive if competition transfers most surplus to customers.

Residual exuberance in Equation \eqref{eq:residexuberance} is useful when applied comparatively. If all firms in a sector have high residuals, the issue may be sector-wide expectations. If residuals are concentrated in firms with high AI narrative exposure but weak cash-flow revisions, bubble risk is more specific. Residuals should be computed under multiple scenarios rather than a single point estimate.

\subsection{SADF and GSADF tests}

The SADF and GSADF procedures identify explosive behavior by estimating recursive right-tailed unit-root tests. \citet{phillips2011} applied explosive-root tests to the NASDAQ run-up, while \citet{phillips2015a,phillips2015b} extended the approach to multiple bubbles and real-time monitoring. For AI, these tests can be applied to individual stocks, industry baskets, valuation ratios, residual-exuberance series, and AI-exposure factor returns.

A stylized regression is
\begin{equation}\label{eq:adf}
\Delta y_t = \alpha + \beta y_{t-1}+\sum_{j=1}^{k}\psi_j \Delta y_{t-j}+\varepsilon_t,
\end{equation}
where a right-tailed test of $\beta>0$ indicates explosive dynamics. The GSADF statistic searches over flexible start and end windows, making it better suited to multiple episodes. For AI, the method's advantage is discipline: it can date-stamp exuberance rather than relying on ex post narratives. Its limitation is that explosive prices can reflect explosive fundamentals, especially in young technologies.

\subsection{LPPL and HLPPL diagnostics}

LPPL models describe bubbles as faster-than-exponential price growth with accelerating oscillations. A standard specification is
\begin{equation}\label{eq:lppl}
\log p(t)=A+B(t_c-t)^m+C(t_c-t)^m\cos\{\omega \log(t_c-t)-\phi\},
\end{equation}
where $t_c$ is a critical time, $m$ controls acceleration, and the cosine term captures log-periodic oscillations. The theory links price acceleration to positive feedback and crash hazard rather than to deterministic crash timing \citep{johansen2000,sornette2001,sornette2003}. Calibration has historically been challenging, and \citet{filimonov2013} propose a more stable transformation that reduces nonlinear complexity.

\citet{cao2025hlppl} extend this tradition by combining LPPL labels, sentiment scores, a hype index, and a dual-stream transformer to produce bubble confidence scores. This is especially relevant to AI because hype is not incidental to AI valuation; it is one of the channels through which expectations, hiring, venture funding, procurement, and media attention interact. HLPPL-style approaches can therefore capture a dimension that price-only LPPL may miss. Their limitations include label construction, backtest overfitting, sector generalization, and the need for transparent validation.

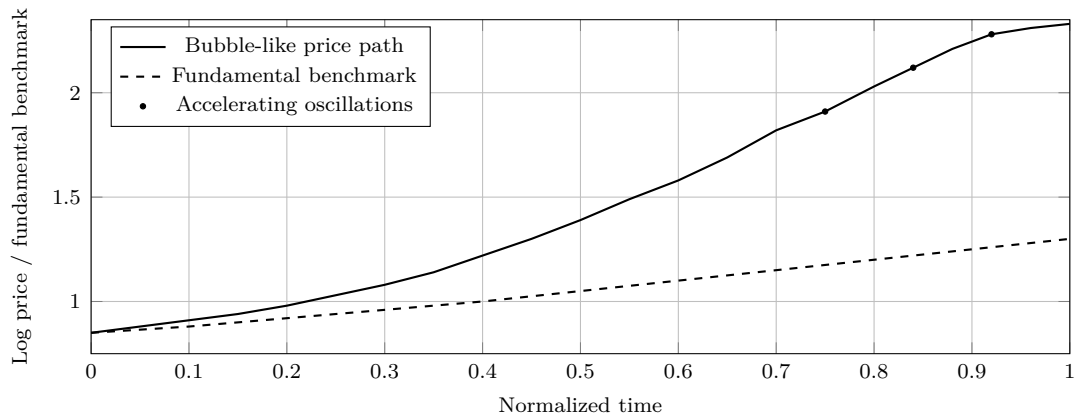
\begin{figure}[H]
\centering
\begin{tikzpicture}
\begin{axis}[
  width=0.88\textwidth,
  height=6cm,
  xlabel={Normalized time},
  ylabel={Log price / fundamental benchmark},
  xmin=0, xmax=1,
  ymin=0.75, ymax=2.35,
  grid=both,
  legend style={at={(0.02,0.98)},anchor=north west,font=\scriptsize},
  ticklabel style={font=\scriptsize},
  label style={font=\scriptsize}
]
\addplot[thick, mark=none] coordinates {
(0.00,0.85) (0.05,0.88) (0.10,0.91) (0.15,0.94) (0.20,0.98) (0.25,1.03) (0.30,1.08) (0.35,1.14) (0.40,1.22) (0.45,1.30) (0.50,1.39) (0.55,1.49) (0.60,1.58) (0.65,1.69) (0.70,1.82) (0.75,1.91) (0.80,2.03) (0.84,2.12) (0.88,2.21) (0.92,2.28) (0.96,2.31) (1.00,2.33)
};
\addlegendentry{Bubble-like price path}
\addplot[thick, dashed, mark=none] coordinates {
(0.00,0.85) (0.10,0.88) (0.20,0.92) (0.30,0.96) (0.40,1.00) (0.50,1.05) (0.60,1.10) (0.70,1.15) (0.80,1.20) (0.90,1.25) (1.00,1.30)
};
\addlegendentry{Fundamental benchmark}
\addplot[only marks, mark=*, mark size=1pt] coordinates {(0.75,1.91) (0.84,2.12) (0.92,2.28)};
\addlegendentry{Accelerating oscillations}
\end{axis}
\end{tikzpicture}
\caption{Stylized bubble-like price path relative to a fundamental benchmark. The figure is conceptual and illustrates faster-than-exponential acceleration and oscillatory behavior often associated with LPPL diagnostics.}\label{fig:lppl}
\end{figure}

\subsection{Machine learning, sentiment, and narrative measurement}

Machine learning can help measure AI exposure, sentiment, and nonlinear interactions. In asset pricing, \citet{gu2020} demonstrate that machine-learning models can improve risk-premium forecasts by capturing nonlinearities. \citet{chen2024deeplearning} use deep learning in an asset-pricing setting with no-arbitrage restrictions. \citet{sirignano2019} identify universal features of price formation using deep learning. These methods suggest that high-dimensional text, fundamentals, and price data can be combined for bubble surveillance.

The challenge is that machine learning can amplify false precision. A model trained on past bubbles may not generalize to AI if AI's economics differ from prior episodes. Textual sentiment may capture attention but not overvaluation. Return predictability may reflect risk premia rather than mispricing. Therefore, machine-learning outputs should be treated as components in an evidence matrix rather than as definitive bubble labels.

\begin{table}[H]
\centering
\caption{Comparison of bubble-detection methods}\label{tab:methods}
\small
\begin{tabularx}{\textwidth}{L{2.6cm}Y Y Y}
\toprule
Method & Strengths & Weaknesses & AI application \\
\midrule
DCF / SDF valuation & Economically interpretable; links price to cash flows and risk & Sensitive to terminal assumptions and omitted growth options & Scenario valuation by AI stack layer \\
Residual exuberance & Separates price from modeled fundamentals & Depends on model quality & Compare AI-exposed firms with similar fundamentals \\
SADF / GSADF & Real-time explosive-dynamics detection; date-stamping & Cannot distinguish explosive fundamentals from bubbles alone & Public AI equities, ETFs, valuation ratios, residuals \\
LPPL & Captures positive feedback and critical-time dynamics & Calibration instability; false positives if poorly constrained & Bubble-like price paths in AI baskets and supplier chains \\
HLPPL / sentiment ML & Integrates hype and market data; suitable for narrative assets & Label and overfitting risk; requires validation & AI narrative intensity, social/media hype, price patterns \\
Issuance and turnover & Historically useful crash predictors & Data may be noisy and sector-specific & IPOs, secondaries, VC rounds, retail turnover \\
Capex-payback tests & Ties valuation to physical investment economics & Requires assumptions on useful life and utilization & Hyperscalers, data centers, power, chip supply chain \\
\bottomrule
\end{tabularx}
\end{table}

\section{A Multi-Pillar Diagnostic Framework}\label{sec:framework}

\begin{figure}[H]
\centering
\begin{tikzpicture}[>=Stealth, node distance=0.75cm and 0.7cm, every node/.style={font=\scriptsize}, diagstep/.style={draw, rounded corners, align=center, text width=2.75cm, minimum height=1.05cm, fill=gray!8}, diagout/.style={draw, rounded corners, align=center, text width=3.2cm, minimum height=1.05cm, fill=gray!12}]
  \node[diagstep] (inputs) {AI exposure map\\stocks, filings, calls, VC, capex};
  \node[diagstep, right=of inputs] (fund) {Fundamental filter\\cash flows, margins, SDF scenarios};
  \node[diagstep, right=of fund] (explosive) {Explosive dynamics\\SADF / GSADF windows};
  \node[diagstep, right=of explosive] (lppl) {LPPL / HLPPL\\critical-time and hype features};
  \node[diagstep, below=of fund] (sent) {Narrative layer\\media, search, analyst revisions, issuance};
  \node[diagstep, below=of explosive] (capex) {Capex sustainability\\useful life, utilization, power, payback};
  \node[diagout, right=of lppl] (score) {Segment-level diagnosis\\Buildout / Boom / Bubble / Bust risk};
  \draw[->, thick] (inputs) -- (fund);
  \draw[->, thick] (fund) -- (explosive);
  \draw[->, thick] (explosive) -- (lppl);
  \draw[->, thick] (lppl) -- (score);
  \draw[->, thick] (inputs.south) |- (sent.west);
  \draw[->, thick] (sent) -- (capex);
  \draw[->, thick] (capex.east) -| (score.south);
  \draw[->, dashed] (sent.north) -- (fund.south);
  \draw[->, dashed] (capex.north) -- (explosive.south);
\end{tikzpicture}
\caption{Multi-pillar diagnostic pipeline for evaluating AI bubble risk. The framework combines fundamentals, explosive dynamics, LPPL/HLPPL patterns, sentiment, and capex sustainability to produce segment-level diagnoses.}\label{fig:pipeline}
\end{figure}
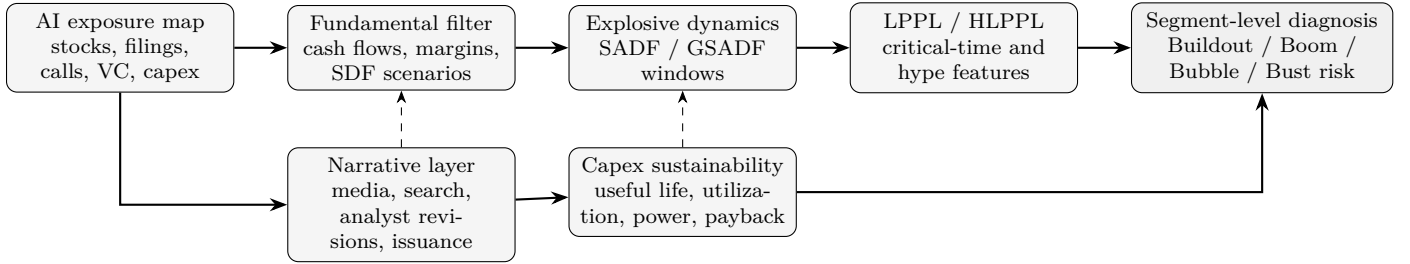

The paper's proposed framework is a five-pillar diagnostic system. It is designed for researchers, investors, and policymakers who need a structured way to evaluate AI-bubble risk without reducing the issue to a slogan. Figure \ref{fig:pipeline} summarizes the pipeline.

\subsection{Pillar 1: AI exposure mapping}

The first step is to define the asset universe and measure AI exposure. Exposure can be measured through business segments, revenue disclosures, capex disclosures, patent and product descriptions, 10-K text, earnings-call transcripts, supplier-customer relationships, and analyst classifications. A firm should not be treated as AI-exposed merely because it uses the term AI in marketing. The framework distinguishes \emph{direct exposure} (AI revenue or AI capex), \emph{input exposure} (supplying chips, power, cooling, or data-center services), \emph{adoption exposure} (using AI to improve operations), and \emph{narrative exposure} (market attention without clear revenue attribution).

\subsection{Pillar 2: Fundamentals and pricing kernels}

The second step estimates fundamentals. For public firms, researchers can use revenue growth, gross margin, operating margin, free cash flow, capex, depreciation, and segment disclosures. For private firms, one must rely on disclosed annualized revenue, funding terms, secondary-market prices, customer contracts, and press reports, with greater uncertainty. The valuation model should include multiple scenarios rather than a single target price. A useful scenario set includes: conservative adoption, baseline adoption, rapid productivity diffusion, commoditization, and infrastructure overbuild.

A pricing-kernel perspective adds discipline. If an asset's payoff is highly correlated with aggregate AI investment and market concentration, it may command a risk premium different from that of a defensive cash-flow asset. Learned SDF models and high-dimensional conditioning variables may help estimate these premia, but their outputs should be interpreted in light of economic restrictions \citep{hansen1991,cochrane2005,gu2020,chen2024deeplearning,chen2026probmeasures}.

\subsection{Pillar 3: Price dynamics}

The third step applies SADF/GSADF and LPPL tests to prices, valuation ratios, and residual-exuberance series. Applying tests to raw prices alone can be misleading in high-growth industries. Applying them to price-to-sales, enterprise-value-to-revenue, or residuals can help isolate exuberance from fundamentals. The framework recommends rolling-window tests for individual assets and stack-level baskets, with robustness checks across currencies, deflators, and benchmark adjustments.

\subsection{Pillar 4: Hype, sentiment, and issuance}

The fourth step measures narrative intensity. Data sources include news volume, earnings-call AI mentions, Google Trends, analyst revisions, social-media attention, ETF flows, IPO volume, secondary offerings, and private funding rounds. The hypothesis is not that hype equals bubble. Hype becomes informative when it rises faster than fundamentals, when it predicts capital raising more than revenue, or when it coincides with deteriorating unit economics. HLPPL-style models are useful because they integrate price feedback and hype in a single architecture \citep{cao2025hlppl}.

\subsection{Pillar 5: Capex sustainability}

\begin{figure}[H]
\centering
\begin{tikzpicture}[>=Stealth, node distance=0.65cm, every node/.style={font=\scriptsize}, box/.style={draw, rounded corners, align=center, text width=3.0cm, minimum height=0.9cm, fill=gray!8}]
  \node[box] (capex) {AI capex\\chips, data centers, power};
  \node[box, right=of capex] (capacity) {Usable compute\\availability, reliability, latency};
  \node[box, right=of capacity] (demand) {Paid demand\\training, inference, enterprise workflows};
  \node[box, right=of demand] (value) {Customer value\\cost savings, revenue uplift};
  \node[box, right=of value] (fcf) {Free cash flow\\margins, retention, payback};
  \foreach \a/\b in {capex/capacity,capacity/demand,demand/value,value/fcf} \draw[->, thick] (\a) -- (\b);
  \node[below=0.7cm of capacity, align=center, text width=3.5cm] (b1) {Bottleneck:\\grid, cooling, supply chain};
  \node[below=0.7cm of demand, align=center, text width=3.3cm] (b2) {Bottleneck:\\model commoditization};
  \node[below=0.7cm of value, align=center, text width=3.3cm] (b3) {Bottleneck:\\adoption and governance};
  \draw[->, dashed] (b1) -- (capacity.south);
  \draw[->, dashed] (b2) -- (demand.south);
  \draw[->, dashed] (b3) -- (value.south);
  \node[above=0.65cm of demand, align=center, text width=6.6cm] {Bubble risk increases when prices capitalize the last box while evidence remains concentrated in the first two boxes.};
\end{tikzpicture}
\caption{Capex-to-monetization funnel for AI infrastructure. The figure is a conceptual synthesis emphasizing that capex is not equivalent to durable free cash flow.}\label{fig:capexfunnel}
\end{figure}
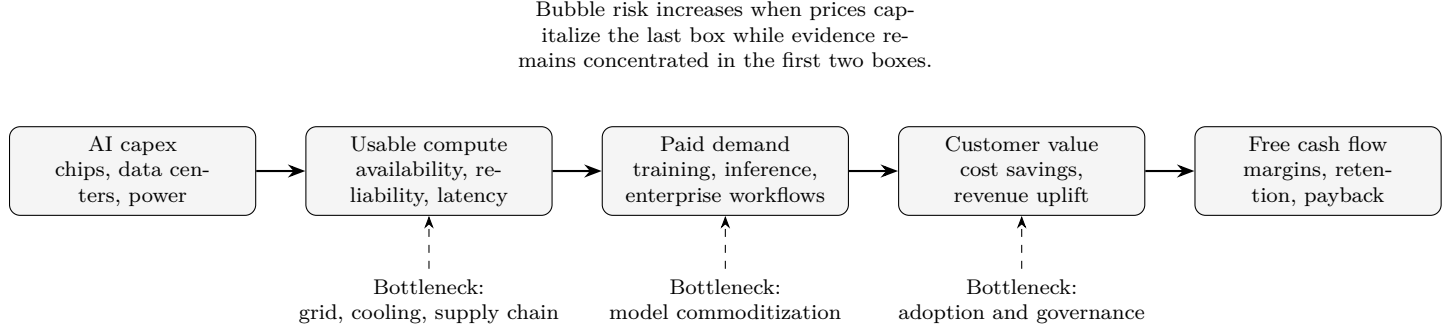

The fifth step evaluates whether AI infrastructure investment can pay for itself. Figure \ref{fig:capexfunnel} summarizes the capex-to-cash-flow funnel. The central question is whether capital expenditure becomes usable compute, whether usable compute becomes paid demand, whether paid demand creates customer value, and whether that value is captured as free cash flow. Bubble risk rises when public valuations capitalize the final free-cash-flow stage while empirical evidence remains concentrated in the initial capex and capacity stages.

The framework produces four segment-level classifications:
\begin{enumerate}[label=(\alph*)]
\item \textbf{Buildout:} large investment with plausible cash-flow support and limited explosive dynamics.
\item \textbf{Boom:} rapid growth with high uncertainty but fundamentals improving in parallel.
\item \textbf{Bubble risk:} prices or valuations accelerate faster than fundamentals, with strong sentiment and weak payback evidence.
\item \textbf{Fragile bubble:} explosive dynamics, extreme residuals, financing dependence, deteriorating unit economics, and high crash-risk indicators.
\end{enumerate}

\section{Illustrative Empirical Design}\label{sec:empirical}

This paper does not invent empirical results. Instead, it specifies a research design that can be implemented using public data and licensed market data. The design is appropriate for a follow-up empirical paper, an institutional research note, or a regulatory surveillance exercise.

\subsection{Research questions and hypotheses}

The main research question is: \emph{Do AI-exposed assets display bubble-like dynamics after controlling for fundamentals, valuation scenarios, sentiment, and capex sustainability?} The proposed hypotheses are:

\begin{description}[leftmargin=1.3cm, style=nextline]
\item[H1: Segmentation.] Bubble evidence is stronger in private AI valuations, application-layer software, and speculative data-center projects than in profitable AI infrastructure leaders.
\item[H2: Capex circularity.] Bubble risk rises when AI capex is justified by future demand from firms whose own revenue depends on continued AI capex elsewhere in the ecosystem.
\item[H3: Fundamental support.] Leading infrastructure firms show weaker bubble signals after controlling for realized revenue growth, margins, and bottleneck rents.
\item[H4: Hype amplification.] Sentiment and issuance variables improve crash-risk prediction beyond price-only explosive-root and LPPL measures.
\item[H5: General-purpose-transition.] AI resembles earlier general-purpose technologies: overvaluation and capital misallocation can coexist with genuine long-run productivity value.
\end{description}

\subsection{Data plan}

Table \ref{tab:data} describes the data plan. The sample would begin before the generative AI boom, ideally in 2018 or earlier, and extend through the latest available data. The universe should include AI-exposed public equities, semiconductor ETFs, software and cloud baskets, data-center REITs, power and grid equipment firms, and broad-market controls. For private firms, the analysis should use funding rounds and disclosed revenue cautiously because valuation terms are often not comparable across rounds.

\begin{table}[H]
\centering
\caption{Data plan and variable definitions}\label{tab:data}
\small
\begin{tabularx}{\textwidth}{L{2.6cm}Y Y Y}
\toprule
Data category & Variables & Sources & Measurement issues \\
\midrule
Public prices & Daily prices, returns, volume, volatility, drawdowns & CRSP, Bloomberg, Refinitiv, exchange data & Survivorship bias, corporate actions, changing index membership \\
Fundamentals & Revenue, gross margin, operating margin, capex, depreciation, free cash flow & Compustat, SEC filings, company reports & AI revenue often not separately disclosed \\
Valuation & P/E, EV/sales, EV/EBITDA, price/book, residual income & Market data and accounting data & Negative earnings and young firms complicate ratios \\
AI exposure & AI mentions, AI segment revenue, product descriptions, supplier-customer links & 10-Ks, earnings calls, transcripts, patents, news & Marketing language may overstate exposure \\
Capex & AI capex, data-center spend, power contracts, GPU purchases & Company filings, IEA, Goldman Sachs, industry reports & Capex attribution and useful life uncertain \\
Sentiment & News volume, search trends, analyst revisions, ETF flows, social attention & News APIs, Google Trends, IBES, ETF flows & Hype can reflect real information or noise \\
Private markets & Funding round size, post-money valuation, revenue, burn rate & PitchBook, CB Insights, company disclosures & Terms, preferences, and revenue quality opaque \\
Bubble diagnostics & SADF/GSADF, LPPL/HLPPL features, residual exuberance & Researcher-constructed & Multiple-testing and calibration risk \\
\bottomrule
\end{tabularx}
\end{table}

\subsection{Empirical strategy}

The empirical strategy has five steps. First, construct AI-exposed portfolios by stack layer. Second, estimate fundamental values or valuation residuals under multiple scenarios. Third, apply SADF/GSADF tests to prices, valuation ratios, and residuals. Fourth, estimate LPPL features and HLPPL-style hype scores. Fifth, test whether the combined indicators predict subsequent drawdowns, volatility, valuation compression, or earnings-revision disappointments.

A baseline panel model for drawdown risk could be written as
\begin{equation}\label{eq:panel}
\Pr(Drawdown_{i,t+h}>d)=\Lambda\left(\alpha_i+\gamma_t+\beta_1 RE_{i,t}+\beta_2 GSADF_{i,t}+\beta_3 LPPL_{i,t}+\beta_4 Hype_{i,t}+\beta_5 PaybackRisk_{i,t}+\mathbf{X}_{i,t}'\delta\right),
\end{equation}
where $\Lambda$ is a logistic link, $\alpha_i$ are firm fixed effects, $\gamma_t$ are time effects, and $\mathbf{X}_{i,t}$ includes controls such as size, profitability, leverage, momentum, volatility, and analyst revisions. The outcome can be defined as a future drawdown greater than threshold $d$, future valuation compression, or a negative earnings-revision event. The model is not intended to prove irrationality; it tests whether bubble-risk indicators have predictive content.

\subsection{Robustness checks}

Robustness checks should include alternative AI-exposure definitions, alternative valuation models, alternative discount rates, sector-neutral portfolios, international samples, market-adjusted returns, deflated prices, and placebo tests on non-AI sectors. The analysis should also test whether signals predict future earnings revisions or only returns. If a signal predicts drawdowns but not fundamentals, it may indicate sentiment reversal; if it predicts earnings disappointments, it may indicate fundamental overestimation.

\begin{table}[H]
\centering
\caption{Research design matrix}\label{tab:researchdesign}
\small
\begin{tabularx}{\textwidth}{L{2.5cm}Y Y Y}
\toprule
Hypothesis & Measurement & Test & Interpretation \\
\midrule
H1 Segmentation & AI stack layer, valuation residuals, bubble scores & Compare indicators across layers & Stronger signals downstream imply localized bubble dynamics \\
H2 Capex circularity & Supplier-customer revenue links, capex dependence, related-party commitments & Network and panel tests & Circular demand raises fragility if cash flows depend on continued funding \\
H3 Fundamental support & Revenue growth, gross margin, free cash flow, backlog & Residual tests after fundamentals & Weak residuals for leaders support buildout interpretation \\
H4 Hype amplification & News, search, AI mentions, issuance, analyst revisions & Incremental predictive power tests & Hype useful if it predicts crashes beyond price dynamics \\
H5 GPT transition & Productivity evidence, adoption, intangible investment, J-curve indicators & Scenario analysis and diffusion models & Coexistence of real value and overvaluation is expected \\
\bottomrule
\end{tabularx}
\end{table}

\section{Current Evidence For and Against an AI Bubble}\label{sec:evidence}

This section summarizes current evidence qualitatively, without presenting new empirical estimates. The evidence is mixed, and the mixed nature is itself informative.

\subsection{Evidence against a pure bubble interpretation}

The first argument against a pure bubble interpretation is realized revenue. Leading AI-infrastructure firms have reported large revenue growth and high margins, which distinguishes parts of the AI boom from pure narrative assets. NVIDIA's fiscal 2027 first-quarter results provide a prominent example of realized AI-infrastructure demand \citep{nvidia2026q1}. A valuation may still be too high, but it is not based solely on imagined revenue.

The second argument is measurable productivity. Task-level studies show that generative AI can increase productivity in writing, customer support, and knowledge-work contexts \citep{noy2023,brynjolfsson2025,dellacqua2023}. These studies are not enough to validate all market valuations, but they show that the technology can produce real output gains.

The third argument is general-purpose technology logic. AI has properties of a GPT: broad applicability, improvement over time, and complementarity with organizational innovation \citep{jovanovic2005,brynjolfsson2021,eloundou2023}. If AI is a GPT, then conventional near-term ratios may understate long-term value in firms that capture durable surplus.

The fourth argument is bottleneck economics. In constrained markets, high margins may reflect scarcity rather than irrationality. Advanced chips, packaging, memory, power interconnection, and data-center capacity are not instantly elastic. Scarcity rents can support high profits until supply catches up.

\subsection{Evidence supporting bubble-risk concerns}

The first concern is capex scale. AI infrastructure spending is enormous relative to observed monetization in some layers. Goldman Sachs' 2026 capex model highlights how sensitive the required buildout is to assumptions about useful life, architecture, and supply constraints \citep{goldman2026tracking}. If useful lives are shorter or utilization lower than expected, returns on invested capital can disappoint.

The second concern is private-market concentration. AI funding has become concentrated in a small number of frontier-model firms and infrastructure providers \citep{cbinsights2026,pitchbook2026}. Concentration is not inherently a bubble, but it increases systemic sensitivity to a small number of valuation marks and funding narratives.

The third concern is monetization uncertainty. Enterprise surveys indicate that adoption is broad but scaled value capture remains uneven \citep{mckinsey2025,pwc2026}. If firms experiment with AI but fail to turn pilots into recurring profits, downstream suppliers may face weaker-than-expected demand.

The fourth concern is narrative intensity. AI is a dominant keyword in corporate communications and investment products. Narrative intensity can be rational when it reflects new information, but it becomes bubble-relevant when capital flows, valuation multiples, and media attention rise faster than earnings revisions. HLPPL-style methods are designed for precisely this interaction between price dynamics and hype \citep{cao2025hlppl}.

The fifth concern is historical base rates. Technology booms often overcapitalize the installation phase. Dot-com history shows that a technology can transform the economy while many investors lose money at peak valuations \citep{ofek2003,greenwood2019}. The same could occur in AI if long-run value accrues to a different set of firms than those receiving current capital.

\begin{table}[H]
\centering
\caption{Evidence matrix for evaluating whether AI is in an ongoing bubble}\label{tab:evidencematrix}
\small
\begin{tabularx}{\textwidth}{L{3.0cm}Y Y}
\toprule
Evidence category & Bubble interpretation & Non-bubble / buildout interpretation \\
\midrule
Revenue growth & Prices may extrapolate temporary scarcity or double ordering & Large realized revenue validates part of the AI infrastructure story \\
Productivity studies & Task-level gains may not scale to aggregate profits & Controlled and field evidence shows genuine productivity potential \\
Capex growth & Investment may outrun monetization and useful life & Front-loaded infrastructure may be necessary for a GPT installation phase \\
Private valuations & Opaque marks and concentrated funding can inflate values & High option value is plausible for platform-scale firms \\
Energy demand & Physical constraints may make projects fragile and expensive & Scarce power and data-center capacity may support durable rents \\
Sentiment and hype & Narrative exposure may drive prices independent of cash flows & High attention may reflect rapid information arrival about a real technology \\
Historical analogies & Dot-com and telecom show overcapitalization risk & Electricity and cloud show delayed but durable productivity gains \\
\bottomrule
\end{tabularx}
\end{table}

The balance of evidence supports a segmented conclusion. The AI boom is not a pure bubble because several core firms have real revenues, productivity evidence exists, and physical bottlenecks create genuine scarcity. But it is not bubble-free because valuations in some layers appear to capitalize optimistic adoption, high margins, long useful lives, and durable pricing power before those assumptions are verified. The framework's main practical recommendation is to classify AI segments by evidence strength rather than to accept or reject the bubble label wholesale.

\section{Financial-Stability and Policy Implications}\label{sec:policy}

A segmented AI bubble would have different financial-stability implications from a broad credit bubble. Household leverage is not the central channel. Instead, potential channels include equity-market concentration, corporate capex cycles, data-center credit exposure, private-market valuation marks, supplier-customer circularity, and energy infrastructure strain.

Equity-market concentration matters because AI leaders have become central to index returns and earnings expectations. If a small set of AI-related firms drives a large share of market capitalization and earnings growth expectations, a valuation shock can affect passive portfolios, retirement accounts, and corporate financing conditions. This does not require AI firms to fail; even slower growth can compress multiples if prices embed extreme expectations.

Corporate capex cycles matter because hyperscaler spending supports revenue across semiconductors, memory, networking, construction, power equipment, and data-center real estate. A capex slowdown could propagate upstream. Conversely, continued capex could crowd out other investment if firms prioritize AI capacity defensively. The macro effect depends on whether AI capex creates productivity-enhancing capital or duplicative capacity.

Private-market marks matter because venture valuations influence employee compensation, secondary transactions, collateral, and fundraising. If private AI valuations are revised downward, the direct financial-system effect may be limited, but the indirect effect on hiring, supplier contracts, and startup demand could be meaningful. The opacity of private terms makes surveillance difficult.

Energy infrastructure creates a policy dimension. AI data centers can accelerate grid investment, renewable procurement, nuclear contracting, and power-market innovation. They can also raise local electricity demand, create permitting conflict, and strain water and land resources. Policy should not assume that all AI-related energy demand is wasteful; nor should it subsidize uneconomic overbuilding. The relevant question is whether pricing, permitting, and grid planning internalize costs and benefits \citep{iea2026energyai}.

The IMF's scenario-planning approach is useful because it treats AI as a macro-financial transition rather than a narrow technology shock \citep{imf2026ai}. Financial regulators should monitor concentration, leverage, interconnectedness, and valuation sensitivity. Corporate boards should require capex payback discipline and scenario analysis. Investors should distinguish firms with realized cash flows and durable moats from firms whose value depends mainly on narrative momentum. Policymakers should support productivity-enhancing diffusion while avoiding policies that amplify speculative financing.

\section{Limitations and Open Problems}\label{sec:limitations}

The framework has several limitations. First, fundamentals are difficult to estimate. AI markets are changing quickly, and public disclosures often do not separate AI revenue from broader cloud, software, or hardware revenue. Private firms disclose even less. Any valuation residual is therefore model-dependent.

Second, bubble tests are not proof of irrationality. Explosive dynamics can reflect explosive fundamentals, and LPPL patterns can be sensitive to sample windows. HLPPL and machine-learning methods add information but introduce model risk, label risk, and overfitting risk. These methods should be used as surveillance tools, not as definitive arbiters.

Third, AI may change the measurement problem itself. If AI improves intangible capital, organizational capability, and consumer surplus, accounting statements may lag value creation. Conversely, if firms capitalize AI narratives without durable margins, accounting may initially overstate economic value through aggressive revenue recognition or related commitments. Distinguishing these cases requires detailed microdata.

Fourth, geopolitical and regulatory factors are central but hard to model. Export controls, chip supply chains, data regulation, copyright litigation, model-safety rules, energy permitting, and antitrust enforcement can shift valuations. These are not exogenous background details; they shape the distribution of AI cash flows.

Fifth, the framework is designed for diagnosis, not deterministic prediction. A market can display bubble risk for years before correcting. It can also deflate through earnings growth rather than price collapse. The right output is a probabilistic segment-level assessment, not a forecast of a crash date.

Table \ref{tab:openproblems} summarizes open research problems.

\begin{table}[H]
\centering
\caption{Open research problems}\label{tab:openproblems}
\small
\begin{tabularx}{\textwidth}{L{3.0cm}Y Y}
\toprule
Problem & Why it matters & Research direction \\
\midrule
AI revenue attribution & Public firms often bundle AI with cloud or software revenue & Textual disclosure models, segment inference, customer-level datasets \\
Private valuation opacity & Terms and preferences make headline valuations incomparable & Contract-level databases, preference-adjusted valuation marks \\
Capex useful life & GPU and data-center economics depend on depreciation speed & Engineering-accounting models of chip generations and utilization \\
Circular financing & Supplier investments may return as customer purchases & Network analysis of strategic investments and purchase commitments \\
Productivity diffusion & Task gains may not scale to aggregate profits & Firm-level adoption panels linked to productivity and margins \\
Hype measurement & Narrative can reflect information or sentiment & Joint models of news, fundamentals, and price dynamics \\
Model commoditization & Open models and inference efficiency can compress margins & Pricing studies across model APIs, enterprise contracts, and open-source adoption \\
Systemic risk & AI concentration may affect equity indices and credit markets & Stress tests for capex slowdown, valuation compression, and supplier shocks \\
\bottomrule
\end{tabularx}
\end{table}

\section{Conclusion}\label{sec:conclusion}

The question of whether AI is in an ongoing bubble cannot be answered responsibly with a simple yes or no. AI is a real technological development with measurable productivity effects, large realized revenues in infrastructure, and plausible general-purpose technology characteristics. At the same time, parts of the AI market display classic bubble-risk features: rapid price appreciation, narrative intensity, concentrated private funding, large capex commitments, uncertain monetization, and assumptions about future productivity that remain only partly verified.

The paper's central conclusion is that AI is best described as a \emph{real technological revolution with localized bubble dynamics}. This conclusion differs from both extreme positions. It rejects the view that AI is merely a speculative mania, because realized cash flows and productivity evidence are meaningful. It also rejects the view that real technology eliminates bubble risk, because history shows that transformative technologies can be overcapitalized during their installation phases.

The proposed diagnostic framework combines five pillars: exposure mapping, fundamental valuation, price-dynamics tests, hype and sentiment measurement, and capex-payback analysis. Its main practical value is segmentation. Leading infrastructure firms with realized revenue and durable bottleneck rents may deserve a buildout or boom classification. Private firms with opaque valuations, application firms with weak moats, and data-center projects dependent on aggressive utilization assumptions may deserve a bubble-risk classification. The classification should update as evidence changes.

For researchers, the AI boom is an opportunity to integrate asset-pricing theory, machine learning, textual analysis, and industrial organization. For investors, it is a reminder that technological truth and investment truth are different: a technology can succeed while many assets disappoint. For policymakers, it is a reason to monitor concentration, financing structures, energy constraints, and the macro-financial consequences of an AI capex cycle. The disciplined question is not whether AI matters. It is where, when, and at what price AI-related cash flows justify the capital now being committed to them.

\newpage
\bibliographystyle{apalike}
\bibliography{references}

\begin{thebibliography}{}

\bibitem[Abreu and Brunnermeier, 2003]{abreu2003}
Abreu, D. and Brunnermeier, M.~K. (2003).
\newblock Bubbles and crashes.
\newblock {\em Econometrica}, 71(1):173--204.

\bibitem[Acemoglu, 2024]{acemoglu2024}
Acemoglu, D. (2024).
\newblock The simple macroeconomics of ai.
\newblock NBER Working Paper 32487, National Bureau of Economic Research.

\bibitem[Acemoglu and Restrepo, 2018]{acemoglu2018}
Acemoglu, D. and Restrepo, P. (2018).
\newblock Artificial intelligence, automation, and work.
\newblock NBER Working Paper 24196, National Bureau of Economic Research.

\bibitem[Agrawal et~al., 2018]{agrawal2018}
Agrawal, A., Gans, J., and Goldfarb, A. (2018).
\newblock {\em Prediction Machines: The Simple Economics of Artificial Intelligence}.
\newblock Harvard Business Review Press, Boston.

\bibitem[Agrawal et~al., 2019]{agrawal2019}
Agrawal, A., Gans, J., and Goldfarb, A., editors (2019).
\newblock {\em The Economics of Artificial Intelligence: An Agenda}.
\newblock University of Chicago Press, Chicago.

\bibitem[Allen and Gale, 2000]{allen2000}
Allen, F. and Gale, D. (2000).
\newblock Bubbles and crises.
\newblock {\em The Economic Journal}, 110(460):236--255.

\bibitem[Arrow, 1953]{arrow1953}
Arrow, K.~J. (1953).
\newblock Le r\^ole des valeurs boursi\`eres pour la r\'epartition la meilleure des risques.
\newblock In {\em Econom\'etrie}, pages 41--47. Centre National de la Recherche Scientifique, Paris.

\bibitem[Bachelier, 1900]{bachelier1900}
Bachelier, L. (1900).
\newblock {\em Th\'eorie de la sp\'eculation}.
\newblock Gauthier-Villars, Paris.
\newblock Originally published in Annales scientifiques de l'\'Ecole Normale Sup\'erieure, 17, 21--86.

\bibitem[Barberis et~al., 1998]{barberis1998}
Barberis, N., Shleifer, A., and Vishny, R. (1998).
\newblock A model of investor sentiment.
\newblock {\em Journal of Financial Economics}, 49(3):307--343.

\bibitem[Barhoumi et~al., 2026]{imf2026ai}
Barhoumi, K., de~Carvalho, F.~A., Gorbanyov, M., Kido, Y., Koll, D., Ostojic, D., Shang, B., Tamirisa, N.~T., Toms, S., Dabla-Norris, E., Nguyen, A. D.~M., and Zhao, Y. (2026).
\newblock Global economic and financial implications of artificial intelligence: Lessons from a scenario planning exercise.

\bibitem[Black and Scholes, 1973]{black1973}
Black, F. and Scholes, M. (1973).
\newblock The pricing of options and corporate liabilities.
\newblock {\em Journal of Political Economy}, 81(3):637--654.

\bibitem[Blanchard and Watson, 1982]{blanchard1982}
Blanchard, O.~J. and Watson, M.~W. (1982).
\newblock Bubbles, rational expectations and financial markets.
\newblock In Wachtel, P., editor, {\em Crises in the Economic and Financial Structure}, pages 295--315. Lexington Books, Lexington, MA.

\bibitem[Breeden and Litzenberger, 1978]{breeden1978}
Breeden, D.~T. and Litzenberger, R.~H. (1978).
\newblock Prices of state-contingent claims implicit in option prices.
\newblock {\em Journal of Business}, 51(4):621--651.

\bibitem[Brunnermeier, 2008]{brunnermeier2008}
Brunnermeier, M.~K. (2008).
\newblock Bubbles.
\newblock {\em The New Palgrave Dictionary of Economics}.

\bibitem[Brynjolfsson et~al., 2025]{brynjolfsson2025}
Brynjolfsson, E., Li, D., and Raymond, L.~R. (2025).
\newblock Generative ai at work.
\newblock {\em The Quarterly Journal of Economics}, 140(2):889--942.

\bibitem[Brynjolfsson et~al., 2021]{brynjolfsson2021}
Brynjolfsson, E., Rock, D., and Syverson, C. (2021).
\newblock The productivity j-curve: How intangibles complement general purpose technologies.
\newblock {\em American Economic Journal: Macroeconomics}, 13(1):333--372.

\bibitem[Cao et~al., 2025]{cao2025hlppl}
Cao, Z., Shao, X., Yan, Y., and Geman, H. (2025).
\newblock Identifying and quantifying financial bubbles with the hyped log-periodic power law model.

\bibitem[{CB Insights}, 2026]{cbinsights2026}
{CB Insights} (2026).
\newblock State of ai 2025 report.
\newblock Research report, January 2026.

\bibitem[Chen, 2026]{chen2026mlfinance}
Chen, K. (2026).
\newblock Machine learning in finance from the first neural-network forecasts to agentic ai: A review summary of applications, evidence, and open problems, 1988--2026.
\newblock Manuscript under review.

\bibitem[Chen et~al., 2024]{chen2024deeplearning}
Chen, L., Pelger, M., and Zhu, J. (2024).
\newblock Deep learning in asset pricing.
\newblock {\em Management Science}, 70(2):714--750.

\bibitem[Cochrane, 2005]{cochrane2005}
Cochrane, J.~H. (2005).
\newblock {\em Asset Pricing}.
\newblock Princeton University Press, Princeton, NJ, revised edition.

\bibitem[David, 1990]{david1990}
David, P.~A. (1990).
\newblock The dynamo and the computer: An historical perspective on the modern productivity paradox.
\newblock {\em American Economic Review Papers and Proceedings}, 80(2):355--361.

\bibitem[Debreu, 1959]{debreu1959}
Debreu, G. (1959).
\newblock {\em Theory of Value: An Axiomatic Analysis of Economic Equilibrium}.
\newblock Yale University Press, New Haven, CT.

\bibitem[Delbaen and Schachermayer, 1994]{delbaen1994}
Delbaen, F. and Schachermayer, W. (1994).
\newblock A general version of the fundamental theorem of asset pricing.
\newblock {\em Mathematische Annalen}, 300:463--520.

\bibitem[Dell'Acqua et~al., 2023]{dellacqua2023}
Dell'Acqua, F., McFowland, E., Mollick, E.~R., Lifshitz-Assaf, H., Kellogg, K., Rajendran, S., Krayer, L., Candelon, F., and Lakhani, K.~R. (2023).
\newblock Navigating the jagged technological frontier: Field experimental evidence of the effects of ai on knowledge worker productivity and quality.
\newblock Working paper, Harvard Business School.

\bibitem[Dixon et~al., 2020]{dixon2020}
Dixon, M.~F., Halperin, I., and Bilokon, P. (2020).
\newblock {\em Machine Learning in Finance: From Theory to Practice}.
\newblock Springer, Cham.

\bibitem[Duffie, 2001]{duffie2001}
Duffie, D. (2001).
\newblock {\em Dynamic Asset Pricing Theory}.
\newblock Princeton University Press, Princeton, NJ, third edition.

\bibitem[Eloundou et~al., 2023]{eloundou2023}
Eloundou, T., Manning, S., Mishkin, P., and Rock, D. (2023).
\newblock Gpts are gpts: An early look at the labor market impact potential of large language models.

\bibitem[Fama, 1970]{fama1970}
Fama, E.~F. (1970).
\newblock Efficient capital markets: A review of theory and empirical work.
\newblock {\em Journal of Finance}, 25(2):383--417.

\bibitem[Fama, 2014]{fama2014}
Fama, E.~F. (2014).
\newblock Two pillars of asset pricing.
\newblock {\em American Economic Review}, 104(6):1467--1485.

\bibitem[Felten et~al., 2021]{felten2021}
Felten, E.~W., Raj, M., and Seamans, R. (2021).
\newblock Occupational, industry, and geographic exposure to artificial intelligence: A novel dataset and its potential uses.
\newblock {\em Strategic Management Journal}, 42(12):2195--2217.

\bibitem[Filimonov and Sornette, 2013]{filimonov2013}
Filimonov, V. and Sornette, D. (2013).
\newblock A stable and robust calibration scheme of the log-periodic power law model.
\newblock {\em Physica A: Statistical Mechanics and its Applications}, 392(17):3698--3707.

\bibitem[Fisher, 1930]{fisher1930}
Fisher, I. (1930).
\newblock {\em The Theory of Interest}.
\newblock Macmillan, New York.

\bibitem[Furman and Seamans, 2019]{furman2019}
Furman, J. and Seamans, R. (2019).
\newblock Ai and the economy.
\newblock {\em Innovation Policy and the Economy}, 19:161--191.

\bibitem[{Goldman Sachs Global Institute}, 2026]{goldman2026tracking}
{Goldman Sachs Global Institute} (2026).
\newblock Tracking trillions: The assumptions shaping the scale of the ai build-out.
\newblock Published May 1, 2026.

\bibitem[Greenwood and Shleifer, 2014]{greenwood2014}
Greenwood, R. and Shleifer, A. (2014).
\newblock Expectations of returns and expected returns.
\newblock {\em Review of Financial Studies}, 27(3):714--746.

\bibitem[Greenwood et~al., 2019]{greenwood2019}
Greenwood, R., Shleifer, A., and You, Y. (2019).
\newblock Bubbles for fama.
\newblock {\em Journal of Financial Economics}, 131(1):20--43.

\bibitem[Grossman and Stiglitz, 1980]{grossman1980}
Grossman, S.~J. and Stiglitz, J.~E. (1980).
\newblock On the impossibility of informationally efficient markets.
\newblock {\em American Economic Review}, 70(3):393--408.

\bibitem[Gu et~al., 2020]{gu2020}
Gu, S., Kelly, B., and Xiu, D. (2020).
\newblock Empirical asset pricing via machine learning.
\newblock {\em Review of Financial Studies}, 33(5):2223--2273.

\bibitem[Hansen and Jagannathan, 1991]{hansen1991}
Hansen, L.~P. and Jagannathan, R. (1991).
\newblock Implications of security market data for models of dynamic economies.
\newblock {\em Journal of Political Economy}, 99(2):225--262.

\bibitem[Harrison and Kreps, 1978]{harrison1978}
Harrison, J.~M. and Kreps, D.~M. (1978).
\newblock Speculative investor behavior in a stock market with heterogeneous expectations.
\newblock {\em Quarterly Journal of Economics}, 92(2):323--336.

\bibitem[Harrison and Kreps, 1979]{harrison1979}
Harrison, J.~M. and Kreps, D.~M. (1979).
\newblock Martingales and arbitrage in multiperiod securities markets.
\newblock {\em Journal of Economic Theory}, 20(3):381--408.

\bibitem[Harrison and Pliska, 1981]{harrison1981}
Harrison, J.~M. and Pliska, S.~R. (1981).
\newblock Martingales and stochastic integrals in the theory of continuous trading.
\newblock {\em Stochastic Processes and their Applications}, 11(3):215--260.

\bibitem[Heaton et~al., 2017]{heaton2017}
Heaton, J.~B., Polson, N.~G., and Witte, J.~H. (2017).
\newblock Deep learning for finance: Deep portfolios.
\newblock {\em Applied Stochastic Models in Business and Industry}, 33(1):3--12.

\bibitem[Hong et~al., 2006]{hong2006}
Hong, H., Scheinkman, J., and Xiong, W. (2006).
\newblock Asset float and speculative bubbles.
\newblock {\em Journal of Finance}, 61(3):1073--1117.

\bibitem[Hutchinson et~al., 1994]{hutchinson1994}
Hutchinson, J.~M., Lo, A.~W., and Poggio, T. (1994).
\newblock A nonparametric approach to pricing and hedging derivative securities via learning networks.
\newblock {\em Journal of Finance}, 49(3):851--889.

\bibitem[{International Energy Agency}, 2026]{iea2026energyai}
{International Energy Agency} (2026).
\newblock Key questions on energy and ai.
\newblock Published April 16, 2026.

\bibitem[Johansen et~al., 2000]{johansen2000}
Johansen, A., Ledoit, O., and Sornette, D. (2000).
\newblock Crashes as critical points.
\newblock {\em International Journal of Theoretical and Applied Finance}, 3(2):219--255.

\bibitem[Jovanovic and Rousseau, 2005]{jovanovic2005}
Jovanovic, B. and Rousseau, P.~L. (2005).
\newblock General purpose technologies.
\newblock In Aghion, P. and Durlauf, S.~N., editors, {\em Handbook of Economic Growth}, volume~1B, pages 1181--1224. Elsevier.

\bibitem[Kindleberger, 1978]{kindleberger1978}
Kindleberger, C.~P. (1978).
\newblock {\em Manias, Panics, and Crashes: A History of Financial Crises}.
\newblock Basic Books, New York.

\bibitem[L\'opez~de Prado, 2018]{lopezdeprado2018}
L\'opez~de Prado, M. (2018).
\newblock {\em Advances in Financial Machine Learning}.
\newblock Wiley, Hoboken, NJ.

\bibitem[{McKinsey QuantumBlack}, 2025]{mckinsey2025}
{McKinsey QuantumBlack} (2025).
\newblock The state of ai: Global survey 2025.
\newblock Global survey report.

\bibitem[Merton, 1973]{merton1973}
Merton, R.~C. (1973).
\newblock Theory of rational option pricing.
\newblock {\em Bell Journal of Economics and Management Science}, 4(1):141--183.

\bibitem[Miller, 1977]{miller1977}
Miller, E.~M. (1977).
\newblock Risk, uncertainty, and divergence of opinion.
\newblock {\em Journal of Finance}, 32(4):1151--1168.

\bibitem[Minsky, 1992]{minsky1992}
Minsky, H.~P. (1992).
\newblock The financial instability hypothesis.
\newblock Working Paper~74, Levy Economics Institute of Bard College.

\bibitem[Noy and Zhang, 2023]{noy2023}
Noy, S. and Zhang, W. (2023).
\newblock Experimental evidence on the productivity effects of generative artificial intelligence.
\newblock {\em Science}, 381(6654):187--192.

\bibitem[{NVIDIA Corporation}, 2026]{nvidia2026q1}
{NVIDIA Corporation} (2026).
\newblock Nvidia announces financial results for first quarter fiscal 2027.
\newblock Press release, May 20, 2026.

\bibitem[Ofek and Richardson, 2003]{ofek2003}
Ofek, E. and Richardson, M. (2003).
\newblock Dotcom mania: The rise and fall of internet stock prices.
\newblock {\em Journal of Finance}, 58(3):1113--1137.

\bibitem[P{\'a}stor and Veronesi, 2009]{pastor2009}
P{\'a}stor, {\v L}. and Veronesi, P. (2009).
\newblock Technological revolutions and stock prices.
\newblock {\em American Economic Review}, 99(4):1451--1483.

\bibitem[Perez, 2002]{perez2002}
Perez, C. (2002).
\newblock {\em Technological Revolutions and Financial Capital: The Dynamics of Bubbles and Golden Ages}.
\newblock Edward Elgar, Cheltenham, UK.

\bibitem[Phillips et~al., 2015a]{phillips2015a}
Phillips, P. C.~B., Shi, S., and Yu, J. (2015a).
\newblock Testing for multiple bubbles: Historical episodes of exuberance and collapse in the s\&p 500.
\newblock {\em International Economic Review}, 56(4):1043--1078.

\bibitem[Phillips et~al., 2015b]{phillips2015b}
Phillips, P. C.~B., Shi, S., and Yu, J. (2015b).
\newblock Testing for multiple bubbles: Limit theory of real-time detectors.
\newblock {\em International Economic Review}, 56(4):1079--1134.

\bibitem[Phillips et~al., 2011]{phillips2011}
Phillips, P. C.~B., Wu, Y., and Yu, J. (2011).
\newblock Explosive behavior in the 1990s nasdaq: When did exuberance escalate asset values?
\newblock {\em International Economic Review}, 52(1):201--226.

\bibitem[{PitchBook}, 2026]{pitchbook2026}
{PitchBook} (2026).
\newblock Q4 2025 ai vc trends.
\newblock Research report, February 2026.

\bibitem[{PwC}, 2026]{pwc2026}
{PwC} (2026).
\newblock 2026 ai performance study.
\newblock Survey report.

\bibitem[Ross, 1976]{ross1976}
Ross, S.~A. (1976).
\newblock The arbitrage theory of capital asset pricing.
\newblock {\em Journal of Economic Theory}, 13(3):341--360.

\bibitem[Samuelson, 1965]{samuelson1965}
Samuelson, P.~A. (1965).
\newblock Proof that properly anticipated prices fluctuate randomly.
\newblock {\em Industrial Management Review}, 6(2):41--49.

\bibitem[Santos and Woodford, 1997]{santos1997}
Santos, M.~S. and Woodford, M. (1997).
\newblock Rational asset pricing bubbles.
\newblock {\em Econometrica}, 65(1):19--57.

\bibitem[Scheinkman and Xiong, 2003]{scheinkman2003}
Scheinkman, J.~A. and Xiong, W. (2003).
\newblock Overconfidence and speculative bubbles.
\newblock {\em Journal of Political Economy}, 111(6):1183--1220.

\bibitem[Schumpeter, 1942]{schumpeter1942}
Schumpeter, J.~A. (1942).
\newblock {\em Capitalism, Socialism and Democracy}.
\newblock Harper and Brothers, New York.

\bibitem[Shiller, 1981]{shiller1981}
Shiller, R.~J. (1981).
\newblock Do stock prices move too much to be justified by subsequent changes in dividends?
\newblock {\em American Economic Review}, 71(3):421--436.

\bibitem[Shiller, 2000]{shiller2000}
Shiller, R.~J. (2000).
\newblock {\em Irrational Exuberance}.
\newblock Princeton University Press, Princeton, NJ.

\bibitem[Shleifer and Vishny, 1997]{shleifer1997}
Shleifer, A. and Vishny, R.~W. (1997).
\newblock The limits of arbitrage.
\newblock {\em Journal of Finance}, 52(1):35--55.

\bibitem[Sirignano and Cont, 2019]{sirignano2019}
Sirignano, J. and Cont, R. (2019).
\newblock Universal features of price formation in financial markets: Perspectives from deep learning.
\newblock {\em Quantitative Finance}, 19(9):1449--1459.

\bibitem[Sornette, 2003]{sornette2003}
Sornette, D. (2003).
\newblock {\em Why Stock Markets Crash: Critical Events in Complex Financial Systems}.
\newblock Princeton University Press, Princeton, NJ.

\bibitem[Sornette and Johansen, 2001]{sornette2001}
Sornette, D. and Johansen, A. (2001).
\newblock Significance of log-periodic precursors to financial crashes.
\newblock {\em Quantitative Finance}, 1(4):452--471.

\bibitem[{Stanford Institute for Human-Centered Artificial Intelligence}, 2026]{stanford2026aiindex}
{Stanford Institute for Human-Centered Artificial Intelligence} (2026).
\newblock The 2026 ai index report.
\newblock Annual report.

\bibitem[Stiglitz, 1990]{stiglitz1990}
Stiglitz, J.~E. (1990).
\newblock Symposium on bubbles.
\newblock {\em Journal of Economic Perspectives}, 4(2):13--18.

\bibitem[Tirole, 1985]{tirole1985}
Tirole, J. (1985).
\newblock Asset bubbles and overlapping generations.
\newblock {\em Econometrica}, 53(6):1499--1528.

\bibitem[White, 1988]{white1988}
White, H. (1988).
\newblock Economic prediction using neural networks: The case of ibm daily stock returns.
\newblock In {\em Proceedings of the Second IEEE Annual Conference on Neural Networks}, volume~2, pages 451--458.

\bibitem[{Zhang Chen} and {Chen Kay}, 2026]{chen2026probmeasures}
{Zhang Chen} and {Chen Kay} (2026).
\newblock Historical developments in probability measures for asset pricing: From state prices to modern pricing kernels.

\end{thebibliography}

\end{document}